\documentclass[twocolumn,prd,aps,groupedaddress,nopacs,nofootinbib]{revtex4}

\usepackage{amsmath}
\usepackage{amssymb}
\usepackage{latexsym}
\usepackage{graphicx}
\usepackage{hyperref}
\usepackage{mathrsfs}
\usepackage{color}
\usepackage{bm}

\begin{document}

\title{Using $H(z)$ data as a probe of the concordance model}

\author{ Marina Seikel$^{1}$, Sahba Yahya$^2$, Roy Maartens$^{2,3}$ and Chris Clarkson$^1$}
\affiliation{ \it $^1$Astrophysics, Cosmology \& Gravity Centre,
and, Department of Mathematics \& Applied Mathematics,
University of Cape Town,
Cape Town 7701,
South Africa
\\
\it $^2$Department of Physics, University of the Western Cape, Cape Town 7535, South Africa\\
\it $^3$Institute of Cosmology \& Gravitation, University of Portsmouth, Portsmouth PO1 3FX, UK
}

\begin{abstract}

Direct observations of the Hubble rate, from cosmic chronometers and
the radial baryon acoustic oscillation scale, can out-perform
supernovae observations in understanding the expansion history,
because supernovae observations need to be differentiated to extract
$H(z)$.  We use existing $H(z)$ data and smooth the data using a new
Gaussian Processes package,
\href{http://www.acgc.uct.ac.za/~seikel/GAPP/index.html}{\sl GaPP},
from which we can also estimate derivatives. The obtained Hubble rate
and its derivatives are used to reconstruct the equation of state of
dark energy and to perform consistency tests of the $\Lambda$CDM
model, some of which are newly devised here. Current data are
consistent with the concordance model, but are rather sparse. Future
observations will provide a dramatic improvement in our ability to
constrain or refute the concordance model of cosmology.  We produce
simulated data to illustrate how effective $H(z)$ data will be in
combination with Gaussian Processes.

\end{abstract}

\maketitle

\section{Introduction}

Different methods and data sets are being used to reconstruct the dark
energy (DE) equation of state $w=p_{\rm de}/\rho_{\rm de}$ and thereby
also to test the concordance model (which has $w=-1$). The results
vary significantly according to the methods and data sets used, and
the error bars and uncertainties are large. It is clear that
higher-precision data are needed for an effective reconstruction and
for robust testing of models. But just as important, more effort is
needed to improve the statistical methods and the design of
observational tests. In particular, there is a need for effective
model-independent statistical methods and for tests that target the
concordance model.

One of the most direct ways to reconstruct $w$ is via supernovae
(SNIa) observations that give the luminosity distance
$d_L$. Model-independent approaches to reconstructing $w$ have been
developed
\cite{weller_albrecht,Alam:2003sc,daly,Alam:2004jy,Wang:2004py,Daly:2004gf,Sahni:2006pa,Shafieloo:2005nd,Zunckel:2007jm,
  EspanaBonet:2008zz,Genovese:2008sw,Bogdanos:2009ib,
  Clarkson:2010bm,Holsclaw:2010sk,Crittenden:2011aa,Shafieloo:2012yh,Lazkoz:2012eh,
  Shafieloo:2012ht,Seikel}.  SNIa observations lead indirectly to
$H(z)$ via the derivative $d_L'(z)$.  Then we need the second
derivative of $d_L(z)$ to reconstruct $w$. This is very challenging
for any reconstruction technique since any noise on the measured
$d_L(z)$ will be magnified in the derivatives.  The problem can be
lessened if direct $H(z)$ data are used because only the first
derivative needs to be calculated to determine $w(z)$.

In this paper we focus on observations that directly give
$H(z)$. Presently, this may be derived from differential ages of
galaxies (`cosmic chronometers') and from the radial baryon acoustic
oscillation (BAO) scale in the galaxy distribution.  Compared to SNIa
observations, less $H(z)$ observational data are needed to reconstruct
$w$ with the same accuracy. For the cosmic chronometer data, it has
been estimated \cite{Ma:2010mr} that 64 data points with the accuracy
of the measurements in \cite{Stern} are needed to achieve the same
reconstruction accuracy as from the Constitution SNIa data
\cite{Hicken}.

We use a model-independent method for smoothing $H(z)$ data to also
perform consistency tests of the concordance model (flat $\Lambda$CDM)
and of curved $\Lambda$CDM models. These consistency tests are
formulated as functions of $H(z)$ and its derivatives which are
constant or zero in $\Lambda$CDM, independently of the parameters of
the model (see~\cite{2012arXiv1204.5505C} for a review). Deviations
from a constant function indicate problems with our assumptions about
dark energy, theory of gravity, or perhaps something else, but without
the usual problems of postulating an alternative to $\Lambda$CDM. Some
of the tests we use here are given for the first time.

Gaussian processes (GP) provide a model-independent smoothing
technique that can meet the challenges of reconstructing derivatives
from data \cite{Rasmussen,MacKay}.  We follow the same GP approach
that has been applied to supernova data in a previous work
\cite{Seikel} by some of the authors of this paper.  We use
\href{http://www.acgc.uct.ac.za/~seikel/GAPP/index.html}{GaPP}
(Gaussian Processes in Python), their publicly available
code\footnote{\href{http://www.acgc.uct.ac.za/~seikel/GAPP/index.html}{\url{http://www.acgc.uct.ac.za/~seikel/GAPP/index.html}}}. (See
\cite{Holsclaw:2010sk,Shafieloo:2012ht} for different uses of GP in
this context.)  A brief description of the GP algorithm is given in
Appendix \ref{GP}.

\begin{figure*}
\includegraphics[width=0.3\textwidth]{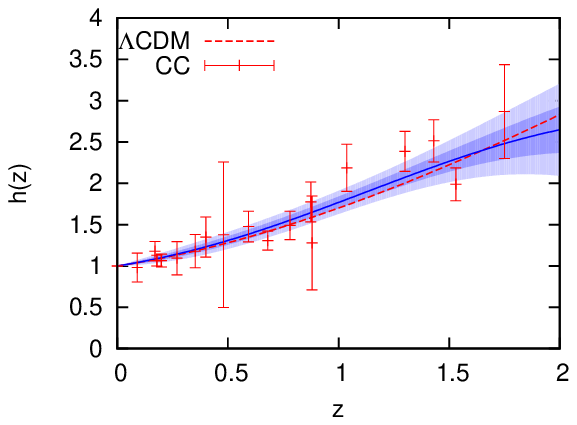}
\includegraphics[width=0.3\textwidth]{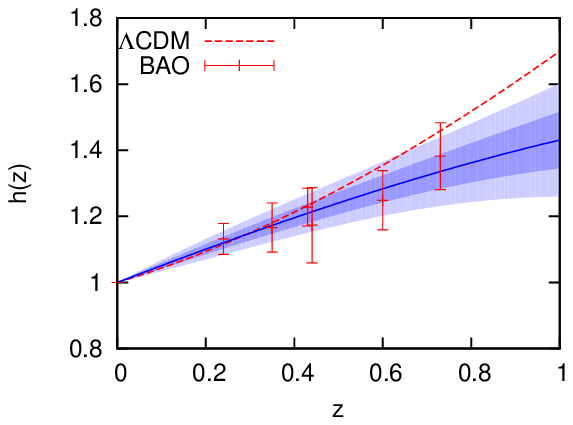}
\includegraphics[width=0.3\textwidth]{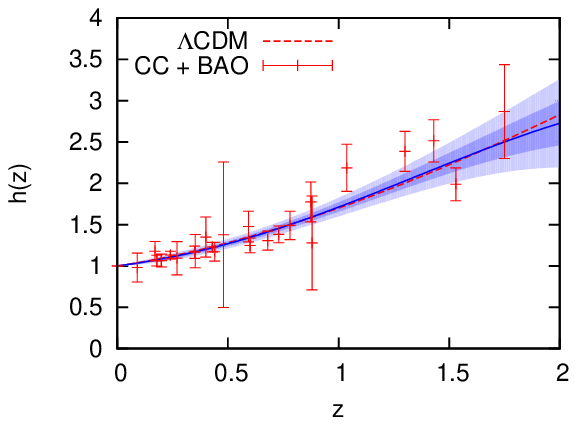}\\
\includegraphics[width=0.3\textwidth]{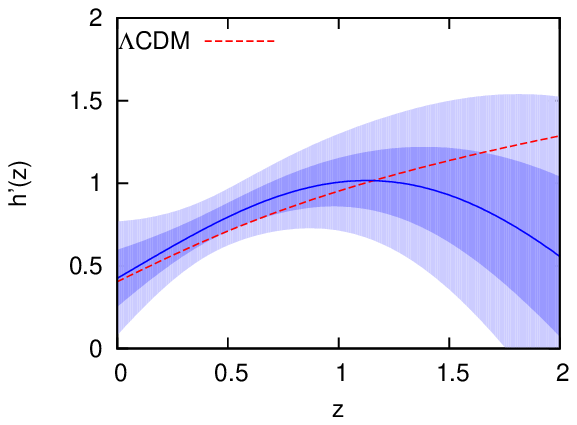}
\includegraphics[width=0.3\textwidth]{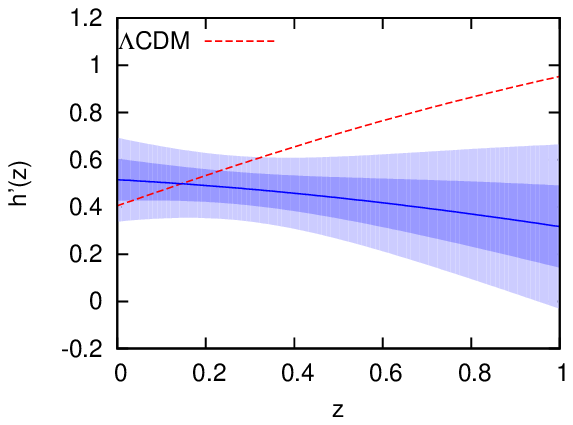}
\includegraphics[width=0.3\textwidth]{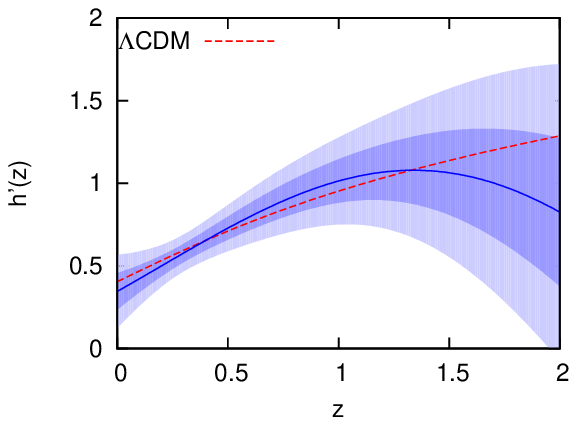}
\caption{$h(z)=H(z)/H_0$ (top) and $h'(z)$ (bottom) reconstructed
  from cosmic chronometer data (left), BAO data (middle) and CC+BAO
  data (right), using Gaussian processes. Shaded areas represent 68\%
  and 95\% confidence levels. The dashed (red) curve is  flat
  $\Lambda$CDM with $\Omega_m =  0.27$; the solid (blue) curve is the
  GP mean. Note that while the BAO data appear to give an inconsistent
  $h'(z)$, this is driven by the two highest redshift points both of
  which happen to lie below the flat $\Lambda$CDM curve.}
\label{hfig}
\end{figure*}

\section{Testing $\Lambda$CDM}\label{theory}

The Friedmann equation,
\begin{eqnarray}
h^2(z) &\equiv& {H^2(z) \over H^2_0}= \Omega_m(1+z)^3 + \Omega_K(1+z)^2 \nonumber\\
&+& (1-\Omega_m - \Omega_K)\exp\left[3 \int_0^z \frac{1+w(z{'})}{1+z{'}} dz{'}\right]\!,
\label{hz}
\end{eqnarray}
can be rearranged to give
\begin{equation}
 w(z)\equiv {p_{\rm de} \over \rho_{\rm de}} =  \frac{ 2(1+z)hh' -
   3h^2 + \Omega_K(1+z)^2}{3\big[h^2 -\Omega_m(1+z)^3 -
     \Omega_K(1+z)^2\big] }\,. 
\label{whz}
\end{equation}
In principle, given $h(z)$ data we can smooth it, attempt to estimate
its derivative, and reconstruct $w(z)$. However, reconstruction of
$w(z)$ is compromised by various difficulties. It depends on the
values of $\Omega_m$ and $\Omega_K$, so we need independent
information about these parameters when we reconstruct $w(z)$ from
$H(z)$ data. These are difficult to estimate without assuming a form
for $w(z)$ \cite{Clarkson:2007bc,Hlozek:2008mt,Kunz:2012aw}.

These difficulties reflect the fact that we cannot use data to
construct physical models~-- rather, we need to use data to test
physical models. The $\Lambda$CDM model could be tested by looking for
deviations from $w=-1$. However, there is a more focused approach: to
develop null hypotheses for $\Lambda$CDM, independently of the
parameters $\Omega_m$ and $\Omega_K$~\cite{2012arXiv1204.5505C}.

To test the concordance model -- i.e. flat $\Lambda$CDM -- we can use
\eqref{hz} to define a  diagnostic function of redshift
\cite{Sahni:2008xx,Zunckel:2008ti,Shafieloo:2009hi}:
\begin{eqnarray}
\mathcal{O}^{(1)}_m(z) &\equiv & \frac{h^2 -1}{z(3+3z+z^2)} . \label{om1h}
\end{eqnarray}
Then
\begin{eqnarray}
\mathcal{O}^{(1)}_m(z) &=&  \Omega_m ~~~\mbox{implies the concordance model}.\nonumber
\end{eqnarray}
If $\mathcal{O}^{(1)}_m(z)$ is not a constant, this is a signal of an
alternative dark energy or modified gravity model. Given observed
$h(z)$ data, we can estimate confidence limits for
$\mathcal{O}^{(1)}_m$. If these are not consistent with a constant
value, we can rule out the concordance model.

It is more effective to measure deviations from zero than from a
constant. The more effective diagnostic is thus the vanishing of the
derivative $\mathcal{O}^{(1)\prime}_m(z)$. This is equivalent to
$\mathcal{L}^{(1)}=0$, where \cite{Zunckel:2008ti}
\begin{eqnarray}
\mathcal{L}^{(1)} &\equiv & 3
   (1+z)^2 (1-h^2)+ 2 z(3+3z+z^2)h h'. \label{Ltest}
\end{eqnarray}
The null test is therefore
\begin{eqnarray}
\mathcal{L}^{(1)}&\neq &0 ~~~\mbox{falsifies the concordance model}. \nonumber
\end{eqnarray}
To apply this test, we need to reconstruct $h'(z)$ from the data.

If the concordance model is ruled out, it is still possible that a
curved $\Lambda$CDM model describes the Universe. 
Equations \eqref{hz} and \eqref{whz} (with $w=-1$) form a linear
system for $\Omega_m$ and $\Omega_K$. Solving for these parameters we
can define 
\begin{eqnarray}
\mathcal{O}^{(2)}_m(z) &\equiv &  2 \frac{ (1+z)(1-h^2)+z(2+z)hh'}{z^2(1+z)(3+z)}\!, \label{Om-hz}\\
\mathcal{O}_K(z) &\equiv & \frac{ 3 (1+z)^2 (h^2 -1)- 2z(3+3z+z^2)hh'}{z^2(1+z)(3+z)}\!, \label{OK-hz}
\end{eqnarray}
and we have
\begin{eqnarray}
\mathcal{O}^{(2)}_m(z)&=&  \Omega_m ~~~\mbox{implies $\Lambda$CDM}, \nonumber\\
\mathcal{O}_K(z) &=&  \Omega_K ~~~\mbox{implies $\Lambda$CDM}. \nonumber
\end{eqnarray}
These quantities are equivalent to those derived in
\cite{Clarkson:2009jq} in terms of $D(z)$, the dimensionless comoving
luminosity distance. The $D(z)$ forms contain second derivatives $D''$
whereas the $h(z)$ forms above contain only first derivatives $h'$.
Given observed Hubble rate data from which we can estimate the
derivative $h'(z)$, we can then estimate confidence limits for ${\cal
  O}^{(2)}_m(z)$ and ${\cal O}^{(2)}_K(z)$. If these are not
consistent with a constant value, we can rule out $\Lambda$CDM in
general, and conclude that dark energy has $w\neq 1$ (or there is
modified gravity).

The more effective diagnostic of these consistency tests is the
vanishing of the derivatives of \eqref{Om-hz} and \eqref{OK-hz}. The
vanishing of $\mathcal{O}^{(2)\prime}_m$ is equivalent to
$\mathcal{L}^{(2)}=0$, where
\begin{eqnarray}
\mathcal{L}^{(2)}(z) &\equiv &  3 (1+z)^2 (h^2- 1)
- 2z(3+6z+2z^2)hh'  \nonumber \\ &+& z^2  (3+z)(1+z)(h'^2+hh'').
\label{Lm2_test}
\end{eqnarray}
Then
\begin{eqnarray}
\mathcal{L}^{(2)}(z)&\neq &0 ~~~\mbox{falsifies $\Lambda$CDM}. \nonumber
\end{eqnarray}
The vanishing of $\mathcal{O}^{(2)\prime}_K$ does not give any
independent information -- it is also equivalent to
$\mathcal{L}^{(2)}=0$.

Given observations of $h(z)$, we can construct this function
independently of the parameters of the model and test $\Lambda$CDM by
measuring consistency with zero. This has the advantage that it is
easier to detect deviations from zero rather than a constant, but at
the expense of requiring an extra derivative in the observable. This
is akin to detecting deviations from constant in $w$, but without
reliance on the parameters of the model. 

For the application of these consistency tests, it is crucial to use a
model-independent method to reconstruct $\mathcal{O}_m^{(1)}$,
$\mathcal{O}_m^{(2)}$, $\mathcal{O}_K$, $\mathcal{L}^{(1)}$ and
$\mathcal{L}^{(2)}$. Model-dependent approaches have the problem that
they affect or even determine the outcome of the consistency test: 
While fitting a $\Lambda$CDM model to the data would always lead to a
result that is consistent with $\Lambda$CDM, fitting a model that does
not include $\Lambda$CDM as a special case would result in
inconsistencies with $\Lambda$CDM. The only model-dependent approches
that do not entirely determine the outcome of the test are those
assuming a model which includes $\Lambda$CDM as a special
case. Nevertheless, they affect the result by forcing the data into a
specific parametrisation, which might not reflect the true model. The
only way to avoid this problem is to use a non-parametric
approach. Here, we use Gaussian processes, which are described in
Appendix \ref{GP}.

\begin{figure*}
\includegraphics[width=0.3\textwidth]{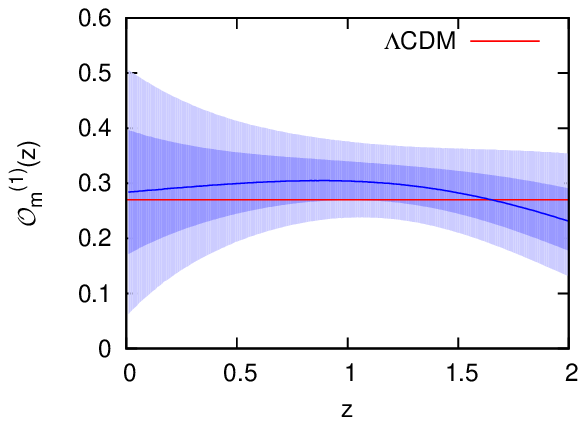}
\includegraphics[width=0.3\textwidth]{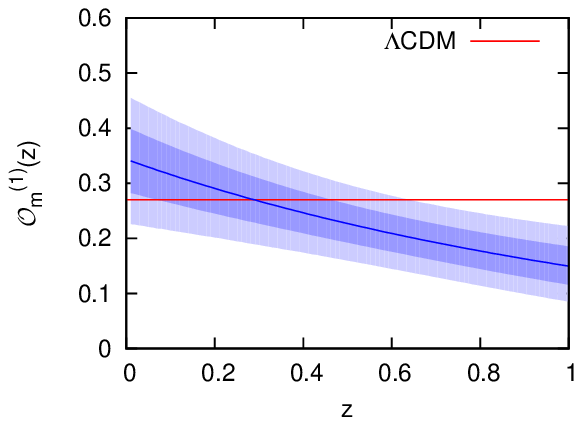}
\includegraphics[width=0.3\textwidth]{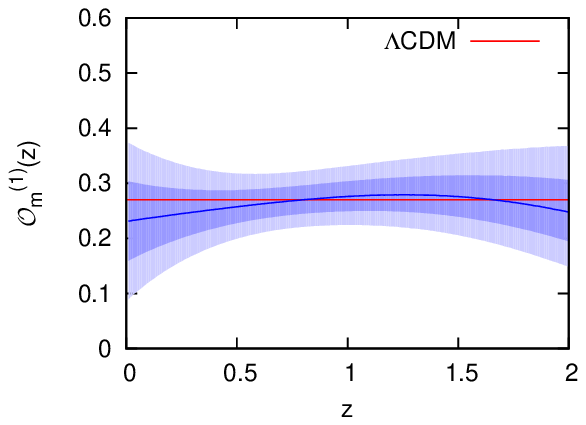}\\
\includegraphics[width=0.3\textwidth]{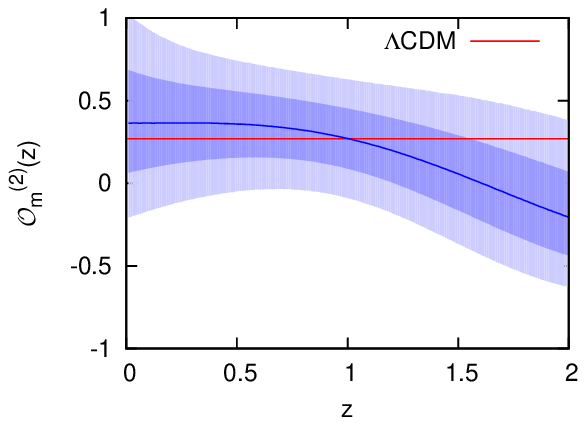}
\includegraphics[width=0.3\textwidth]{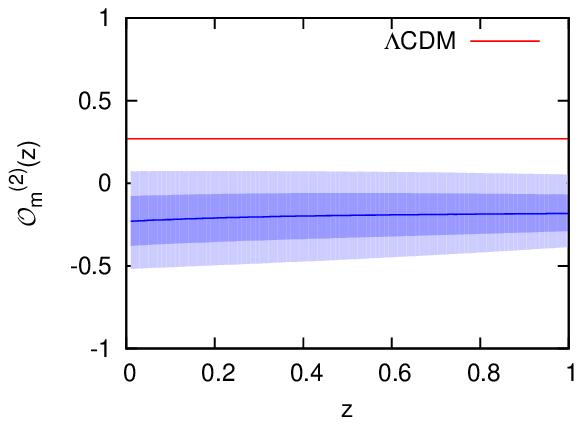}
\includegraphics[width=0.3\textwidth]{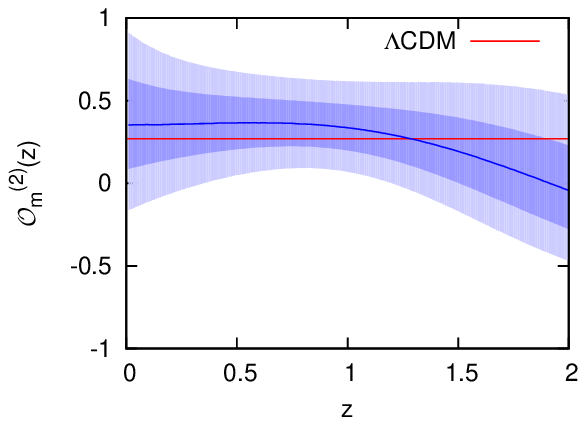}\\
\includegraphics[width=0.3\textwidth]{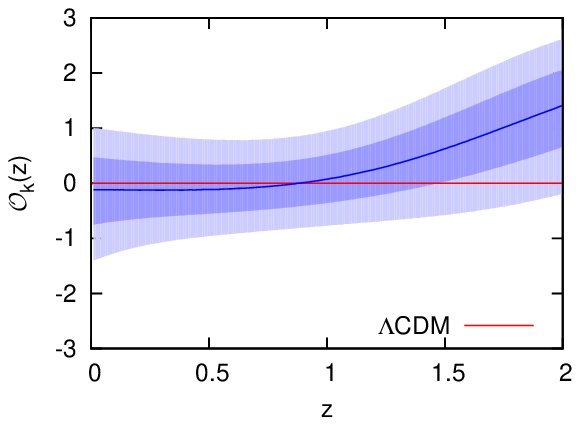}
\includegraphics[width=0.3\textwidth]{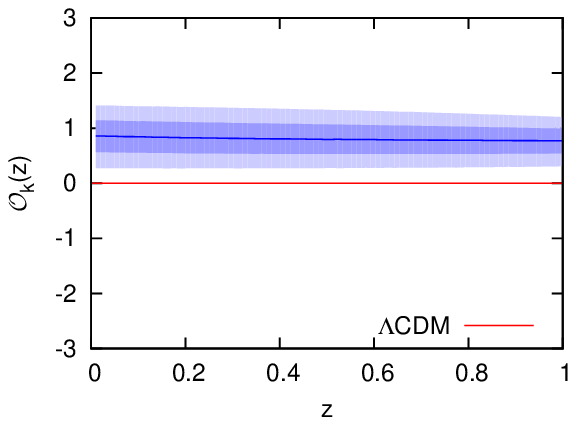}
\includegraphics[width=0.3\textwidth]{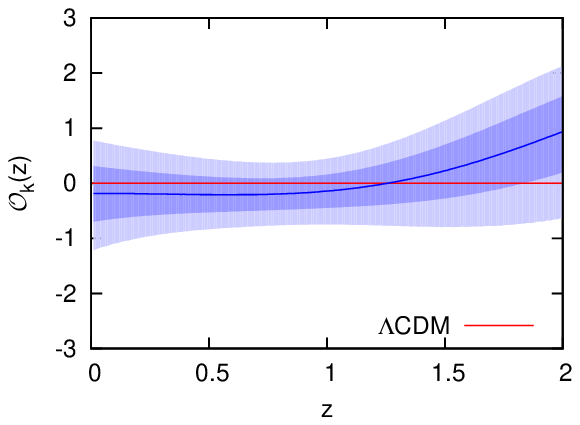}
\caption{$\mathcal{O}^{(1)}_m(z)$ (top), $\mathcal{O}^{(2)}_m(z)$
  (middle) and $\mathcal{O}_K(z)$ (bottom) reconstructed from
  cosmic chronometers (left), BAO (middle) and CC+BAO (right).
  For $\mathcal{O}^{(1)}_m(z)$, the dashed (red) curve is flat
  $\Lambda$CDM. For $\mathcal{O}^{(2)}_m(z)$ and
  $\mathcal{O}_K(z)$ it is a curved $\Lambda$CDM model.}
\label{Om1}\label{Om2}\label{Ok}
\end{figure*}

\section{Reconstruction and consistency tests from $H(z)$ data}\label{reconstruction}

Cosmic chronometers are based on observations of the differential ages
of galaxies
\cite{Stern,Jimenez:2001gg,Crawford:2010rg,Moresco:2012jh}. The Hubble
rate at an emitter with redshift $z$ is
\begin{equation}
H(z) =  -\frac{1}{1+z} \frac{dz}{dt_e},
\end{equation}
where $t_e$ is the proper time of emission. The differential method
uses passively evolving galaxies formed at the same time to determine
the age difference $\Delta t_e$ in a small redshift bin $\Delta z$,
assuming a Friedmann background. To find old galaxies sharing the same
formation time, we have to look for the oldest stars in both galaxies
and show that they have the same age. This method is effective; but
while the differential approach significantly
reduces the systematics that would be present when determining the
absolute ages of galaxies, it
still faces uncertainties due to the assumptions that are made to
estimate the age.

The second way to measure $H(z)$ is the observed line-of-sight
redshift separation $\Delta z$ of the baryonic acoustic oscillation
(BAO) feature in the galaxy 2-point correlation function
\cite{Gaztanaga:2008xz,Chuang,Blake:2012pj},
\begin{equation}
H(z) = \frac{\Delta z}{r_s(z_d)}\,,
\end{equation}
where $r_s(z_d)$ is the sound horizon at the baryon drag epoch.

\begin{figure*}
\includegraphics[width=0.3\textwidth]{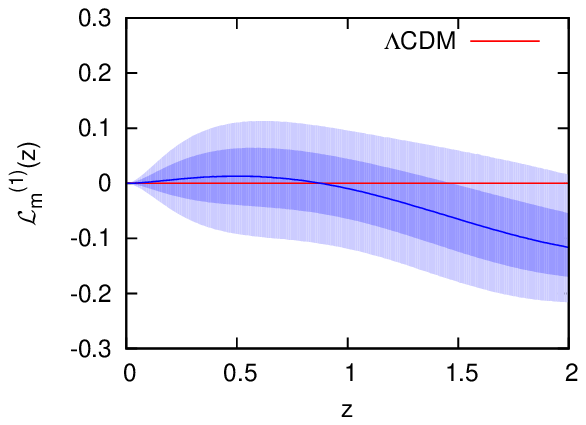}
\includegraphics[width=0.3\textwidth]{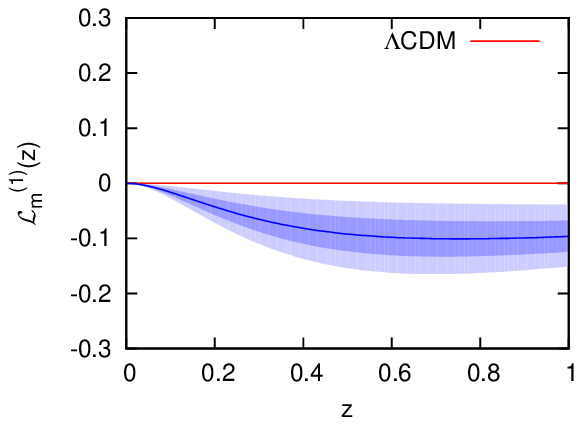}
\includegraphics[width=0.3\textwidth]{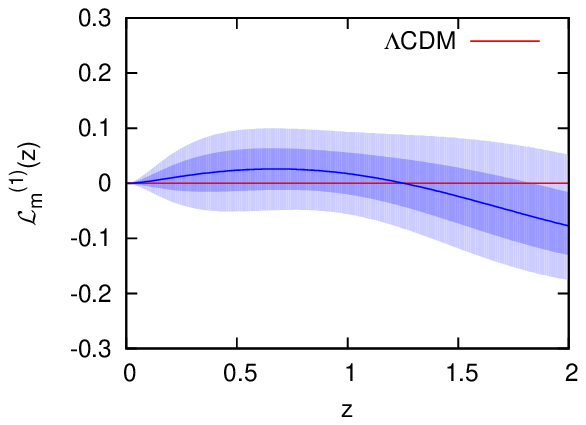}\\
\includegraphics[width=0.3\textwidth]{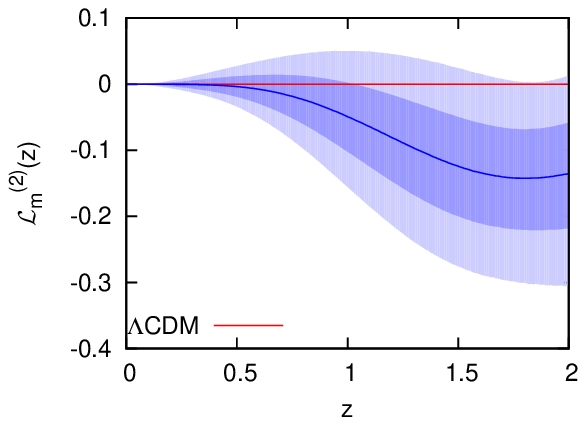}
\includegraphics[width=0.3\textwidth]{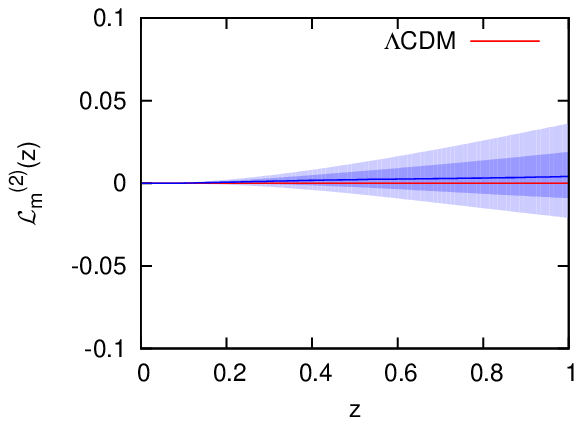}
\includegraphics[width=0.3\textwidth]{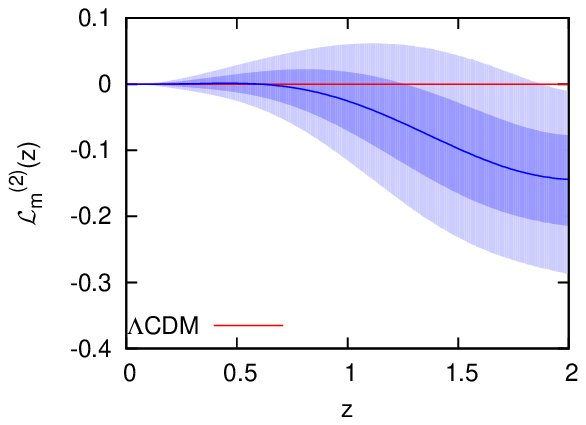}
\caption{$\mathcal{L}^{(1)}_m=\mathcal{L}^{(1)}/(1+z)^6$ (top) and
  $\mathcal{L}^{(2)}_m=\mathcal{L}^{(2)}/(1+z)^6$ 
  (bottom) reconstructed from  cosmic chronometers (left), BAO (middle) and CC+BAO
  (right). The dashed (red) curve is a $\Lambda$CDM model. }
\label{Lm1_Lm2}
\end{figure*}
\begin{figure*}
\includegraphics[width=0.3\textwidth]{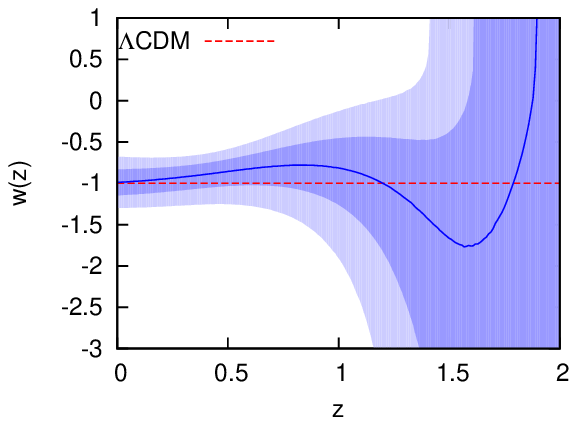}
\includegraphics[width=0.3\textwidth]{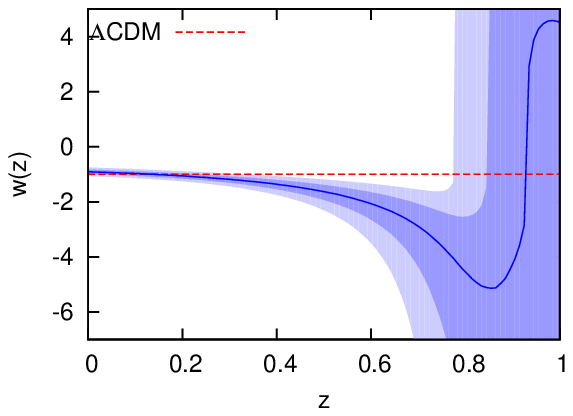}
\includegraphics[width=0.3\textwidth]{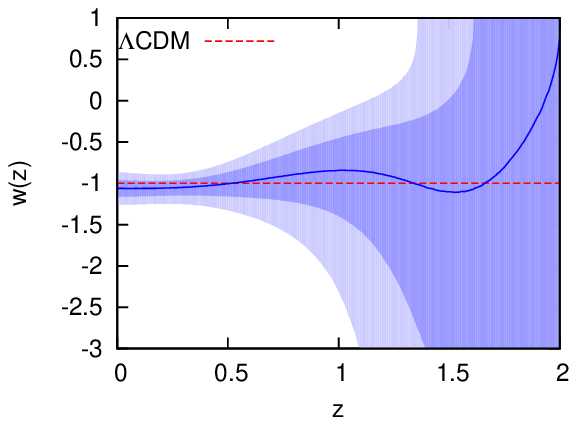}
\caption{$w(z)$ reconstructed from  cosmic chronometers (left), BAO
  (middle~-- note the different $z$ range) and CC+BAO
  (right) by marginalizing over $\Omega_m = 0.275 \pm 0.016$.
  The dashed (red) curve is a $\Lambda$CDM model.}
\label{wfig}
\end{figure*}

\subsection*{Results: real data}

\noindent We use the following $H(z)$ data sets:\\[1mm]
{\em CC:} ~~18 cosmic chronometer data points  \cite{Moresco:2012by}.\\
{\em BAO:} ~~6 radial BAO data points \cite{Blake:2012pj,Gaztanaga:2008xz,Chuang}.\\
{\em CC+BAO:} ~~Combination of CC and BAO sets.\hspace{4mm}

We normalize $H(z)$ using $H_0 = 70.4 \pm
2.5\,$km\,s${}^{-1}$Mpc${}^{-1}$. The uncertainty in $H_0$ is
transferred to $h(z)$ as $ \sigma_h^2 = ({\sigma_H^2}/{H_0^2}) +
({H^2}/{H_0^4}) \sigma^2_{H_0}.$ The reconstructed functions $h(z)$
and $h'(z)$ are shown in Fig.~\ref{hfig}. The shaded regions
correspond to the 68\% and 95\% confidence levels (CL). The true model
is expected to lie 68\% of the plotted redshift range within the 68\%
CL. Note that this is only an expectation value. The actual value for
a specific function may deviate from the expectation. The dependence
of the actual percentage on the smoothness of the function has been
analysed in \cite{Seikel}.

Figure \ref{Om1} shows the reconstruction of $\mathcal{O}_m^{(1)}$.
The reconstruction of $\mathcal{O}_m^{(2)}$ and $\mathcal{O}_K$ is
shown in Fig. \ref{Om2}, and Fig.~\ref{Lm1_Lm2} gives
$\mathcal{L}^{(1)}$ and $\mathcal{L}^{(2)}$. We actually plot a
modified $\mathcal{L}_m=\mathcal{L}/(1+z)^6$ which stabilises the
errors at high redshift without affecting the consistency condition.
The reconstructed $w(z)$, also requiring $h'$, is shown in
Fig. \ref{wfig}, where we assume the concordance values $\Omega_m =
0.275 \pm 0.016$ and $\Omega_K= 0$ \cite{Komatsu}.
\begin{figure*}
\includegraphics[width=0.3\textwidth]{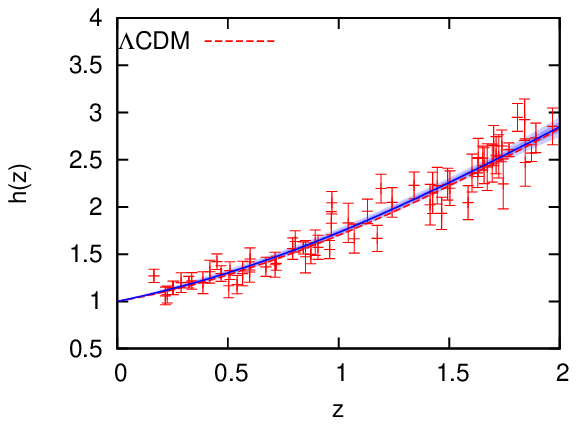}
\includegraphics[width=0.3\textwidth]{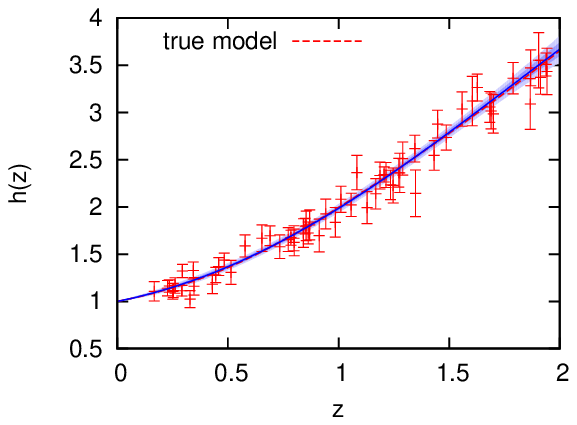}\\
\includegraphics[width=0.3\textwidth]{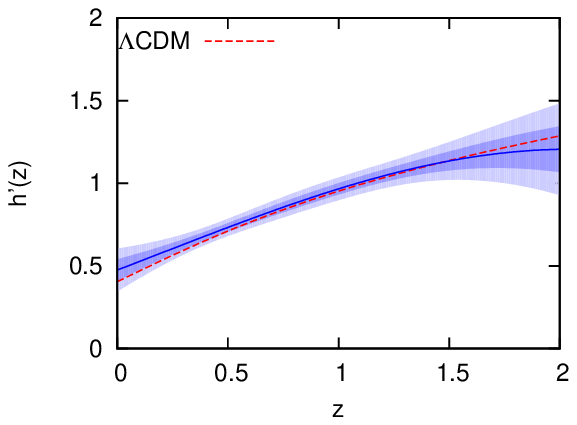}
\includegraphics[width=0.3\textwidth]{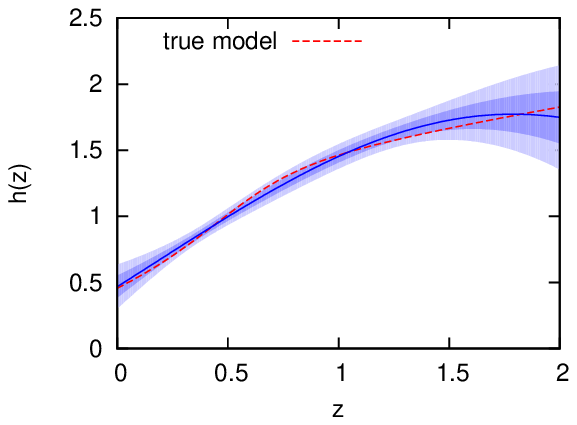}\\
\includegraphics[width=0.3\textwidth]{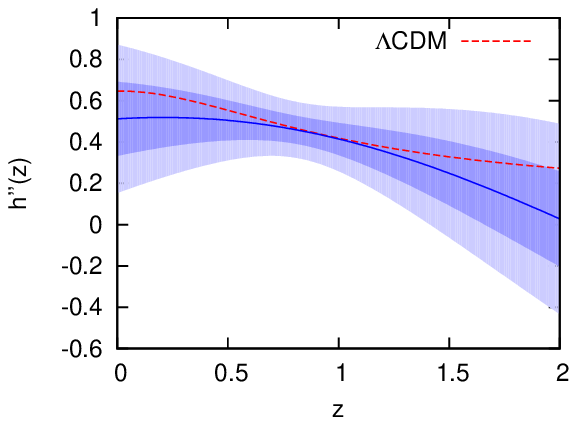}
\includegraphics[width=0.3\textwidth]{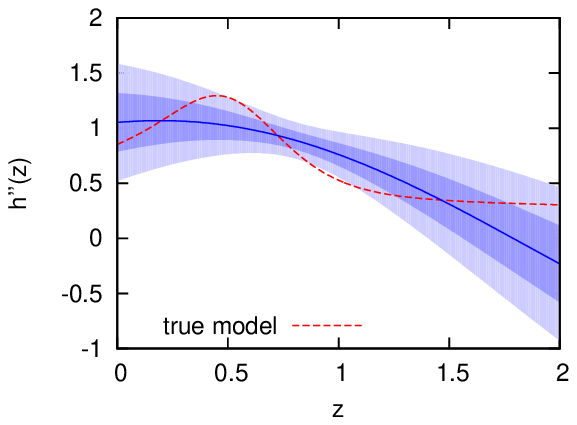}
\caption{$h(z)$ (top), $h'(z)$ (middle) and $h''(z)$ (bottom)
  reconstructed from
  simulated data, assuming a concordance model (left) and model \eqref{wzevo} with slowly
  evolving $w(z)$ (right). }
\label{hmock}
\end{figure*}
\begin{figure*}
\includegraphics[width=0.3\textwidth]{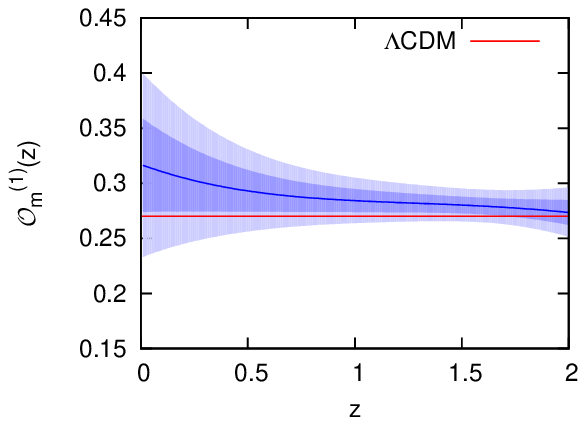}
\includegraphics[width=0.3\textwidth]{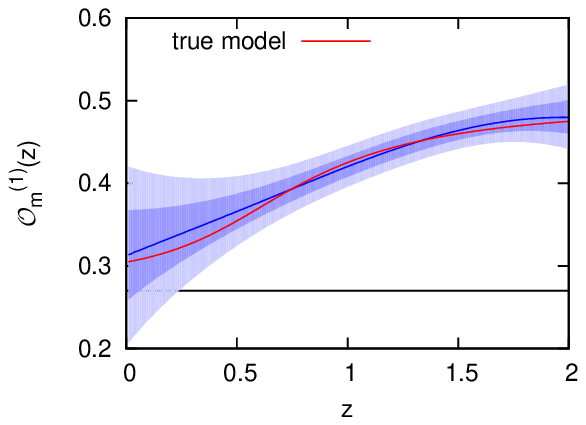}\\
\includegraphics[width=0.3\textwidth]{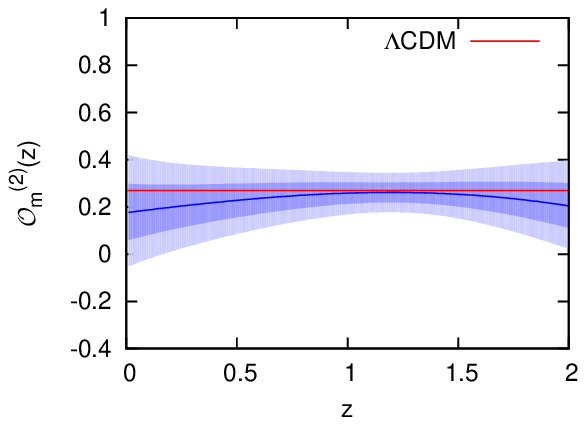}
\includegraphics[width=0.3\textwidth]{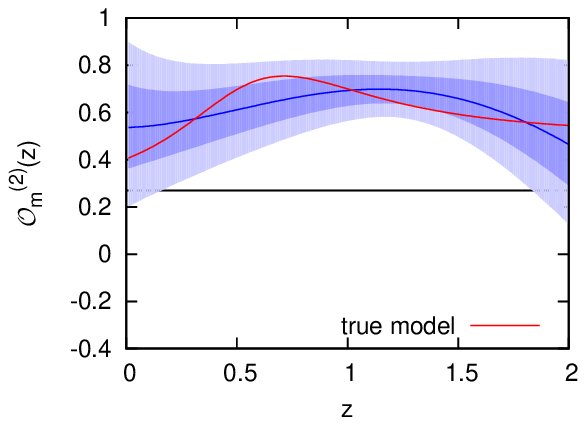}\\
\includegraphics[width=0.3\textwidth]{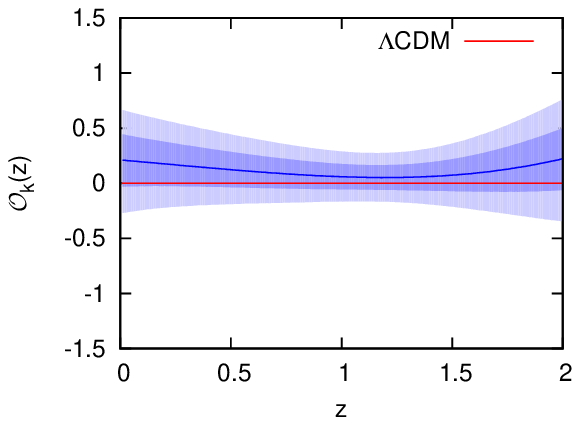}
\includegraphics[width=0.3\textwidth]{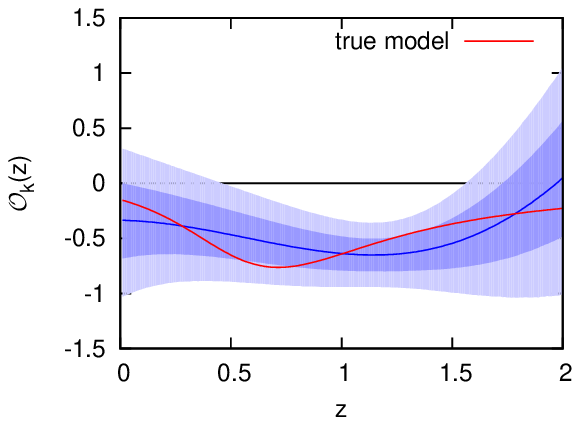}
\caption{$\mathcal{O}^{(1)}_m(z)$ (top), $\mathcal{O}^{(2)}_m(z)$
  (middle) and $\mathcal{O}_K(z)$ (bottom) reconstructed from
  simulated data, assuming a concordance model (left) and model \eqref{wzevo} (right).}
\label{mock_OmOk}
\end{figure*}
\begin{figure*}
\includegraphics[width=0.3\textwidth]{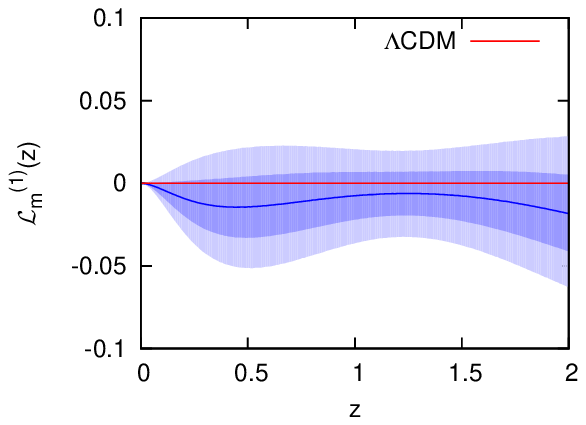}
\includegraphics[width=0.3\textwidth]{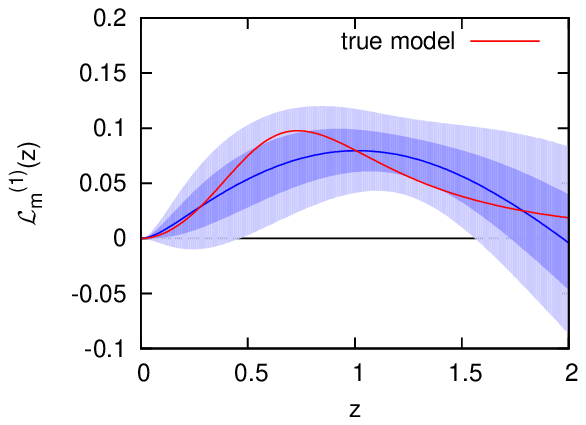}\\
\includegraphics[width=0.3\textwidth]{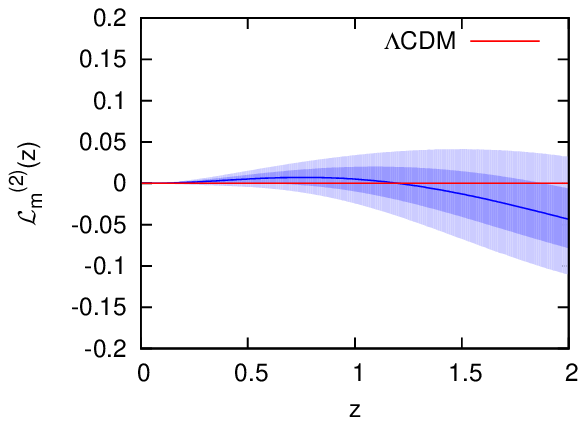}
\includegraphics[width=0.3\textwidth]{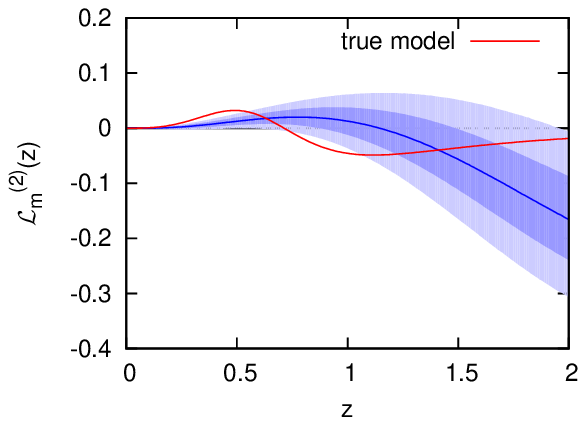}
\caption{$\mathcal{L}^{(1)}_m=\mathcal{L}^{(1)}/(1+z)^6$ (top) and
  $\mathcal{L}^{(2)}_m=\mathcal{L}^{(2)}/(1+z)^6$ 
  (bottom) reconstructed from
  simulated data, assuming a concordance model (left) and model \eqref{wzevo} (right). }
\label{mock_Lm}
\end{figure*}
\begin{figure*}
\includegraphics[width=0.4\textwidth]{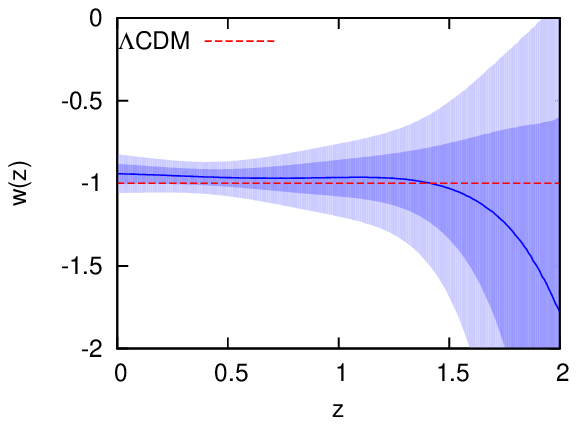}
\includegraphics[width=0.4\textwidth]{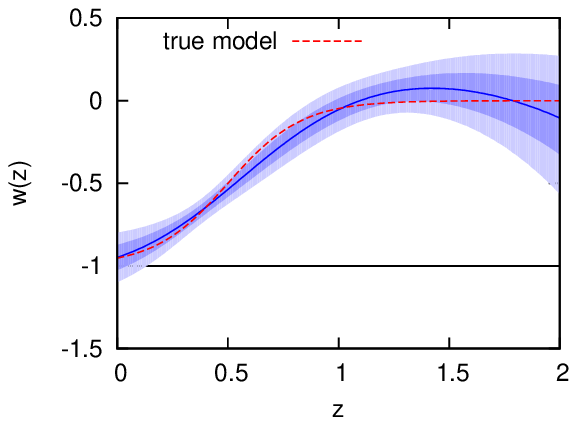}
\caption{$w(z)$ reconstructed from
  simulated data, assuming a concordance model (left) and model \eqref{wzevo} (right), by marginalizing over $\Omega_m = 0.275 \pm 0.016$.}
\label{mockw}
\end{figure*}

\subsection*{Results: mock data}

To demonstrate how a larger number of data will affect our results
when reconstructing $w$ and testing $\Lambda$CDM, we simulated a data
set of 64 points for $H(z)$, drawing the error from a Gaussian
distribution $\mathcal{N}(\bar{\sigma},\epsilon)$ with $\bar{\sigma} =
10.64z+8.86$ and $\epsilon = 0.125(12.46z+3.23)$, adapting the
methodology of~\cite{Ma:2010mr}.

We simulated data points for two different models:\\
Concordance model, $\Omega_K = 0$, $\Omega_m = 0.27$.\\
A model with slowly evolving equation of state:
\begin{equation}
w(z) =  -\frac{1}{2}
 +{1\over2} \tanh3 \Big(z- \frac{1}{2}\Big),
\label{wzevo}
\end{equation}
and the same concordance density parameters.

The GP reconstructions are shown in
Figs. \ref{hmock}--\ref{mockw}.

\subsection*{Discussion}\label{conclusion}

Figure \ref{Om1} shows that for the CC and CC+BAO data (18 and 24
points), we get good reconstructions when there is no differentiation
of $h(z)$ involved.  The BAO data set only contains 6 data points up
to redshift 0.73. Beyond that redshift, the reconstruction differs
significantly from $\Lambda$CDM. The results from the CC and CC+BAO
sets are however in very good agreement with $\Lambda$CDM.

The BAO data appear to be inconsistent with the concordance
model. However, 6 data points are not sufficient for a reliable
reconstruction. The two data points with highest redshift happen to be
below the concordance curve, which pulls the reconstructed curve
down. This is probably just a coincidence, but it illustrates the
importance of having the derivative of the data consistent with the
model, as well as the data itself. Current and upcoming large-volume
surveys, such as BOSS~\cite{Schlegel:2009hj}, EUCLID~\cite{euclid}
and SKA~\cite{Ansari:2011bv}, will provide radial BAO measurements of
increasing number and precision.

The reconstruction of $\mathcal{O}_m^{(2)}$ and $\mathcal{O}_K$ shown
in Fig. \ref{Om2} is more challenging for the available data set,
since we need the first derivative of $h$.  With present data sets,
the uncertainties in the reconstruction are quite large. Using CC and
CC+BAO, these results as well as the results for $\mathcal{L}^{(1)}$
and $\mathcal{L}^{(2)}$ shown in Fig.~\ref{Lm1_Lm2}, are consistent
with $\Lambda$CDM.

For the mock data sets, Figs.~\ref{hmock} and \ref{mock_OmOk} show
that the GP reconstructions recovers the assumed models very
effectively. We can clearly distinguish the model with slowly evolving
$w(z)$ from $\Lambda$CDM in $\mathcal{O}_m^{(1)}$. For
$\mathcal{O}_m^{(2)}$ and $\mathcal{O}_K$, the reconstruction errors
are too large to see this difference.  The same is true for
consistency tests $\mathcal{L}^{(1)}$ and $\mathcal{L}^{(2)}$ shown in
Fig.~\ref{mock_Lm}.

The reconstruction of the equation of state $w(z)$ also shows a clear
difference of the two models, assuming we can accurately determine
$H_0$, $\Omega_m$ and $\Omega_K$ separately from $w(z)$: see
Fig.~\ref{mockw}. GP works very well to recover the assumed $w$.  With
less than 100 data points, we can reconstruct a dynamical dark energy
model far better than is achievable using thousands of SNIa data~--
compare to analogous reconstructions in~\cite{Seikel}.

\section{Conclusions}

We have considered the information that current and future $H(z)$ data
can give us. Currently such data come from cosmic chronometers and BAO
data, and is plainly consistent with the concordance model. Future
data, however, will provide a powerful discriminator between models.
It is remarkable how few data points are required compared to
supernovae: to reconstruct $w(z)$ accurately in our non-parametric way
requires many thousands of SNIa, compared to less than 100 $H(z)$ data
points.

We have derived and analysed new consistency tests for the
$\Lambda$CDM model, which we have formulated in terms of $H(z)$
directly, rather than using the more familiar distance
function~\cite{Clarkson:2009jq,2012arXiv1204.5505C}. By smoothing the
data points using Gaussian process, we have shown that these can be
very effective in determining that $\Lambda$CDM is the incorrect
model, but without having to assume the key parameters $\Omega_m$ and
$\Omega_K$, which currently only have constraints derived by assuming
$\Lambda$CDM or a similar alternative. These tests not only require
that the data points themselves are consistent with the model, but
that their derivative is also.

Future data which directly measures the expansion history will
therefore play an important role in future dark energy studies.

~\\{\bf Acknowledgements:}\\
We thank Phil Bull and Mat Smith for discussions. 
SY and RM are supported by the South African Square Kilometre Array
Project. MS and CC are supported by the National Research Foundation
(NRF) South Africa. RM is supported by the UK Science \& Technology
Facilities Council (grant no. ST/H002774/1). 

\appendix

\section{Gaussian Processes}\label{GP}

For a data set $ \{(z_i,y_i)| i = 1, \dots ,n\}$, where $\bm Z$
represents the training points $z_i$, i.e. the locations of the
observations, we want to reconstruct the function that describes the
data at the test input points $\bm Z^*$.

A Gaussian Process is a distribution over functions and is thus a
generalization of a Gaussian distribution. It is defined by the mean
$\mu(z)$ and covariance $k(z, \tilde{z})$:
\begin{equation}
 f(z) \sim \mathcal{GP}\left(\mu(z), k(z,\tilde{z})\right)\,.
\end{equation}
At each $z_i$, the value $f(z_i)$ is drawn from a
Gaussian distribution with mean $\mu(z_i)$ and variance
$k(z_i,z_i)$. $f(z_i)$ and $f(z_j)$
are correlated by the covariance function $k(z_i,z_j)$.

Choosing the covariance function is one of the main points for
achieving satisfactory results. The squared exponential is a general
purpose covariance function, which we use throughout this paper:
\begin{equation}
 k(z_i,z_j) = \sigma_f^2 \exp\left[-\frac{(z_i -z_j)^2}{2 \ell^2}\right]\,.
\label{cov}
\end{equation}
The `hyperparameters' are $\sigma_f$ (signal variance) and $\ell$
(characteristic length scale). $\ell$ can be thought of as the
distance moved in input space before the function value changes
significantly. $\sigma_f$ describes the typical change in
$y$-direction. In contrast to actual parameters, they do not specify
the exact form of a function, but describe typical changes in the
function value.

For $\bm Z^*$, the covariance matrix is given by $[K(\bm Z^*,\bm
  Z^*)]_{ij}=k(z^*_i,z^*_j)$. Then the vector $\bm f^*$ with entries
$f(z^*_i)$ is drawn from a Gaussian distribution:
\begin{equation}\label{prior}
\bm f^* \sim \mathcal{N}(\bm\mu(\bm Z^*), K(\bm Z^*, \bm Z^*))\,.
\end{equation}
This can be considered as a prior for the distribution of $\bm f^*$. One
needs to add observational information to obtain the posterior distribution.

The observational data have a covariance matrix $C$. For uncorrelated
data, $C$ is a diagonal matrix with entries $\sigma_i$. The combined
distribution for $\bm f^*$ and the observations $\bm y$ is given by:
\begin{equation}
\begin{bmatrix} {\bm y} \\ {\bm f^*} \end{bmatrix} \sim
\mathcal{N} \left(
\begin{bmatrix}
\text{\boldmath $\mu$}\\
\text{\boldmath $\mu^*$}
\end{bmatrix}
,
\begin{bmatrix}
K(\bm Z,\bm Z) + C & K(\bm Z,\bm Z^*)\\
K(\bm Z^*,\bm Z)   & K(\bm Z^*,\bm Z^*)
\end{bmatrix}
\right)
\end{equation}

While the values of $\bm y$ are already known, we want to reconstruct
$\bm f^*$. Thus, we are interested in the conditional distribution
\begin{equation}\label{post}
{\bm f^*}| \bm Z^*, \bm Z,\bm y \sim \mathcal{N} \left(
\bar{\bm f}^*, \text{cov}({\bm f^*})
\right) \;,
\end{equation}
where
\begin{eqnarray}
&&\bar{\bm f}^* = \text{\boldmath $\mu^*$} + K(\bm Z^*,\bm Z)\left[K(\bm Z,\bm Z) +
  C\right]^{-1}\! ({\bm y} -  \text{\boldmath $\mu$})~~~~~\\
&&\text{cov}({\bm f^*})=  K(\bm Z^*,\bm Z^*) \nonumber\\
&&{}- K(\bm Z^*,\bm Z) \left[K(\bm Z,\bm Z) +
  C\right]^{-1}  K(\bm Z,\bm Z^*),
\end{eqnarray}
are the mean and covariance of ${\bm f^*}$, respectively. The variance
of ${\bm f^*}$ is simply the diagonal of $\text{cov}({\bm
  f^*})$. Equation (\ref{post}) is the posterior distribution of the
function given the data and the prior (\ref{prior}).

In order to use this equation, we need to know the values of the
hyperparameters $\sigma_f$ and $\ell$. They can be trained by
maximizing the log marginal likelihood:
\begin{eqnarray}
\ln \mathcal{L} &=& \ln p({\bm y}|\bm Z,\sigma_f,\ell) \nonumber\\
&=& -\frac{1}{2}({\bm
  y} - \text{\boldmath $\mu$})^{\top} \left[K(\bm Z,\bm Z) +
  C\right]^{-1} ({\bm y} - \text{\boldmath
  $\mu$}) \nonumber\\
&& {} - \frac{1}{2}\ln
  \left|K(\bm Z,\bm Z)+C\right|
  - \frac{n}{2}\ln 2\pi  \;.
\end{eqnarray}
Note that this likelihood only depends on the observational data, but
is independent of the locations $\bm Z^*$ where the function is to be
reconstructed.

Derivatives of the function can be reconstructed in a similar way. For
the first derivative, the conditional distribution is given by \cite{Seikel}:
\begin{equation}
{\bm f^*}'| \bm Z^*, \bm Z,y \sim \mathcal{N} \left(
\bar{{\bm f}}^*{'}, \text{cov}({\bm f^*}')
\right) \;,
\end{equation}
where
\begin{eqnarray}
&&\bar{{\bm f}}^*{'} = \text{\boldmath $\mu^*$}' + K'(\bm Z^*,\bm Z)
\left[K(\bm Z,\bm Z) + C\right]^{-1}\! ({\bm y}-\text{\boldmath $\mu$})~~~~~~~~\\
&&\text{cov}({\bm f^*}') = K''(\bm Z^*,\bm Z^*)\nonumber\\ &&{}- K'(\bm Z^*,\bm Z)\left[K(\bm
  Z,\bm Z) + C \right]^{-1}  K'(\bm Z,\bm Z^*).
\end{eqnarray}
For the covariance matrices, we use the notation:
\begin{eqnarray}
[K'(\bm Z,\bm Z^*)]_{ij} & =& \frac{\partial k(z_i,z^*_j)}{\partial z^*_j}\\
{}[K''(\bm Z^*,\bm Z^*)]_{ij} & = & \frac{\partial^2k(z^*_i,z^*_j)}{\partial
    z^*_i\,\partial z^*_j} \;.
\end{eqnarray}
$K'(\bm Z^*,\bm Z)$ is the transpose of $K'(\bm Z,\bm Z^*)$.

To calculate a function $g(f,f')$ which depends on $f$
and $f'$, we also need to know the covariances between $f^*=f(z^*)$
and $f^*{}'=f'(z^*)$ at each point $z^*$ where $g$ is to be
reconstructed. This covariance is given by:
\begin{eqnarray}
\text{cov}(f^*,f^*{}') &=&\left.\frac{\partial k(z^*,\tilde{z})}{\partial
  \tilde{z}}\right|_{z^*} \\ &-& K'(z^*,\bm Z)
\left[K(\bm Z,\bm Z)+C\right]^{-1} K(\bm Z,z^*).\nonumber
\end{eqnarray}
$g^*=g(z^*)$ is then determined by Monte Carlo sampling, where in each
step $f^*$ and $f^*{}'$ are drawn from a multivariate normal distribution:
\begin{equation}
\begin{bmatrix} {f^*} \\ {f^*{}'} \end{bmatrix} \sim
\mathcal{N} \left(
\begin{bmatrix}
\bar{f^*}\\
\bar{f^*{}'}
\end{bmatrix}
,
\begin{bmatrix}
\text{var}(f^*)       & \text{cov}(f^*,f^*{}') \\
\text{cov}(f^*,f^*{}')  & \text{var}(f^*{}')
\end{bmatrix}
\right) \,.
\end{equation}


\begin{thebibliography}{99}



\bibitem{weller_albrecht}
 J.~Weller and A.~Albrecht,
  Phys.\ Rev.\ D {\bf 65}, 103512 (2002)

\bibitem{Alam:2003sc}
  U.~Alam, V.~Sahni, T.~D.~Saini and A.~A.~Starobinsky,
 MNRAS  {\bf 344}, 1057 (2003)

\bibitem{daly}
  R.~A.~Daly and S.~G.~Djorgovski,
  Astrophys.\ J.\  {\bf 597}, 9 (2003)


\bibitem{Alam:2004jy}
  U.~Alam, V.~Sahni and A.~A.~Starobinsky,
  JCAP {\bf 0406}, 008 (2004)

\bibitem{Wang:2004py}
  Y.~Wang and M.~Tegmark,
  Phys.\ Rev.\ Lett.\  {\bf 92}, 241302 (2004)

\bibitem{Daly:2004gf}
  R.~A.~Daly and S.~G.~Djorgovski,
  AJ  {\bf 612}, 652 (2004)

  


\bibitem{Shafieloo:2005nd}
  A.~Shafieloo, U.~Alam, V.~Sahni and A.~A.~Starobinsky,
  Mon.\ Not.\ Roy.\ Astron.\ Soc.\  {\bf 366}, 1081 (2006)


\bibitem{Sahni:2006pa}
  V.~Sahni and A.~Starobinsky,
  Int.\ J.\ Mod.\ Phys.\  D {\bf 15}, 2105 (2006)




\bibitem{Zunckel:2007jm}
  C.~Zunckel and R.~Trotta,
  Mon.\ Not.\ Roy.\ Astron.\ Soc.\  {\bf 380}, 865 (2007)


\bibitem{EspanaBonet:2008zz}
  C.~Espana-Bonet and P.~Ruiz-Lapuente,
  JCAP {\bf 0802}, 018 (2008)

\bibitem{Genovese:2008sw}
  C.~R.~Genovese, P.~Freeman, L.~Wasserman, R.~C.~Nichol and C.~Miller,
  arXiv:0805.4136.

\bibitem{Bogdanos:2009ib}
  C.~Bogdanos and S.~Nesseris,
  JCAP {\bf 0905}, 006 (2009)

\bibitem{Clarkson:2010bm}
  C.~Clarkson and C.~Zunckel,
  Phys.\ Rev.\ Lett.\  {\bf 104}, 211301 (2010)

\bibitem{Holsclaw:2010sk}
  T.~Holsclaw, U.~Alam, B.~Sanso, H.~Lee, K.~Heitmann, S.~Habib and D.~Higdon,
  Phys.\ Rev.\ Lett.\  {\bf 105}, 241302 (2010)

\bibitem{Crittenden:2011aa}
  R.~G.~Crittenden, G.~-B.~Zhao, L.~Pogosian, L.~Samushia and X.~Zhang,
  JCAP {\bf 1202}, 048 (2012)

\bibitem{Shafieloo:2012yh}
  A.~Shafieloo,
  arXiv:1204.1109.


\bibitem{Lazkoz:2012eh}
  R.~Lazkoz, V.~Salzano and I.~Sendra,
  arXiv:1202.4689 [astro-ph.CO].



\bibitem{Shafieloo:2012ht}
  A.~Shafieloo, A.~G.~Kim and E.~V.~Linder,
  arXiv:1204.2272.

\bibitem{Seikel}
M.~Seikel, C.~Clarkson and M.~Smith, arXiv:1204.2832.

\bibitem{Ma:2010mr}
  C.~Ma and T.~-J.~Zhang,
  Astrophys.\ J.\  {\bf 730}, 74 (2011)

\bibitem{Stern}
  D.~Stern, R.~Jimenez, L.~Verde, M.~Kamionkowski and S.~A.~Stanford,
  JCAP {\bf 1002}, 008 (2010)

\bibitem{Hicken}
  M. Hicken {\em et al.},
  Astrophys. J. {\bf 700}, 1097 (2009)

  \bibitem[Clarkson(2012)]{2012arXiv1204.5505C} C. Clarkson, 
arXiv:1204.5505. 


\bibitem{Rasmussen}
C.~Rasmussen and C.~Williams, {\em Gaussian Processes for Machine Learning}
(MIT Press, 2006).

\bibitem{MacKay}
D.~MacKay, {\em Information Theory, Inference and Learning
  Algorithms} (Cambridge University Press, 2003).

\bibitem{Clarkson:2007bc} 
  C.~Clarkson, M.~Cortes and B.~A.~Bassett,
  JCAP {\bf 0708}, 011 (2007)

\bibitem{Hlozek:2008mt}
  R.~Hlozek, M.~Cortes, C.~Clarkson and B.~Bassett,
  arXiv:0801.3847.

\bibitem{Kunz:2012aw} 
  M.~Kunz,
  arXiv:1204.5482 [astro-ph.CO].

\bibitem{Zunckel:2008ti}
  C.~Zunckel and C.~Clarkson,
  Phys.\ Rev.\ Lett.\  {\bf 101}, 181301 (2008)

\bibitem{Sahni:2008xx}
  V.~Sahni, A.~Shafieloo and A.~A.~Starobinsky,
  Phys.\ Rev.\ D {\bf 78} (2008) 103502

\bibitem{Shafieloo:2009hi}
  A.~Shafieloo and C.~Clarkson,
  Phys.\ Rev.\ D {\bf 81}, 083537 (2010)

\bibitem{Clarkson:2009jq}
  C.~Clarkson,
  AIP Conf.\ Proc.\  {\bf 1241}, 784 (2010)

\bibitem{Jimenez:2001gg}
  R.~Jimenez and A.~Loeb,
  Astrophys.\ J.\  {\bf 573}, 37 (2002)

\bibitem{Crawford:2010rg}
  S.~M.~Crawford, A.~L.~Ratsimbazafy, C.~M.~Cress, E.~A.~Olivier, S-L.~Blyth and K.~J.~van der Heyden,
  arXiv:1004.2378.

\bibitem{Moresco:2012jh}
  M.~Moresco, A.~Cimatti, R.~Jimenez, L.~Pozzetti, G.~Zamorani, M.~Bolzonella, J.~Dunlop and F.~Lamareille {\it et al.},
  arXiv:1201.3609.

\bibitem{Moresco:2012by}
  M.~Moresco, L.~Verde, L.~Pozzetti, R.~Jimenez and A.~Cimatti,
  arXiv:1201.6658.

\bibitem{Gaztanaga:2008xz}
  E.~Gaztanaga, A.~Cabre and L.~Hui,
  Mon.\ Not.\ Roy.\ Astron.\ Soc.\  {\bf 399}, 1663 (2009)

\bibitem{Chuang}
  C.-H.~Chuang and Y.~Wang,
  arXiv:1102.2251.

\bibitem{Blake:2012pj}
  C.~Blake, S.~Brough, M.~Colless, C.~Contreras, W.~Couch, S.~Croom, D.~Croton and T.~Davis {\it et al.},
  arXiv:1204.3674.

\bibitem{Komatsu}
E. Komatsu {\em et al.}, Astrophys. J. Suppl. {\bf 192}, 18 (2011) [arXiv:1001.4538].


\bibitem{Schlegel:2009hj}
  D.~Schlegel {\it et al.},
  arXiv:0902.4680.
  
\bibitem{euclid}
W. Percival,  {\tt sci.esa.int/science-e/www/object/ doc.cfm?fobjectid=46450}

\bibitem{Ansari:2011bv}
  R.~Ansari, J.~E.~Campagne, P.~Colom, J.~M.~L.~Goff, C.~Magneville, J.~M.~Martin, M.~Moniez and J.~Rich {\it et al.},
  arXiv:1108.1474 [astro-ph.CO].

 \end{thebibliography}
\end{document}